\documentclass{aa}
\usepackage{graphicx}
\begin{document}


\title{Gravitational instability of expanding shells}
\subtitle{Solution with nonlinear terms}
\author{R. W\"unsch \and J. Palou\v s}

\institute{Astronomical Institute, Academy of Sciences of the Czech Republic
, Bo\v cn\'\i\ II 1401, 141 31 Praha 4, Czech Republic}

\date{Received 9 October 2000 / Revised 11 April 2001 / Accepted 29 May 2001}

\abstract{
A model of the thin shell expanding into a uniform ambient medium
is developed.
Density perturbations are described using equations with linear
and quadratic terms, and the linear and the nonlinear solutions are 
compared.
We follow the time evolution of the fragmentation process and separate 
the well
defined fragments. Their mass spectrum is compared to observations and we 
also estimate their  formation time.  
}
\maketitle

\section{Introduction}
$\mathrm{H_{\alpha}}$ emission, which is frequently observed along the periphery of 
giant expanding HI shells, is related to star formation, which has been 
triggered due to the compression of the ISM. As an example of 
correlated star formation, the ring of young stars observed near 
to the Sun consisting of Orion, Perseus, Sco-Cen and other OB associations
can be used (Lindblad et al., 1997).
The HI hole No. 35 in the dwarf galaxy IC 2574 (Walter \& Brinks, 1999) 
with the $\mathrm{H_{\alpha }}$ 
emission found along its rims serves as  another example of 
triggered star formation.

The analysis of the stability of a pre\-ssure-con\-fined 
slab performed by Elmegreen \& Elmegreen~(\cite{elmegreen2}) 
has been extended to spherical
expanding shocks by Vishniac~(\cite{vish1}). The time development
of the gravitational collapse of linear perturbations in decelerating,
isothermal shocked layers has been examined numerically and analytically
by Elmegreen~(\cite{elmegreen1}, \cite{elmegreen}). The purely hydrodynamical
nonlinear instability has been discussed by Vishniac~(\cite{vish2}).
In this paper, we continue with nonlinear analysis of the gravitational 
instability of spherically symmetric shells expanding into stationary 
homogeneous medium.

We modify the approach adopted by Fuchs~(\cite{fuchs})
who described the fragmentation of uniformly rotating self-gravitating 
disks. If some conditions are fulfilled, the expanding shell may become 
gravitationally unstable and break to fragments.
The inclusion of higher order terms helps to determine with better 
accuracy than the linear analysis when, where and how quickly it happens.

The Rayleigh-Taylor (R-T) instability is not expected to develop in the 
situation explored because a spherical shock expanding into the homogeneous
interstellar medium is always decelerated collecting stationary ambient 
medium. R-T instabilities may be important in different situations when the 
density of the ambient medium drops down sufficiently quickly so that the
shell can accelerate mixing hot and cold gas components. This is the
case of very active SF regions where the shell interacts with previously
formed fragments. The region 30 Doradus in LMC may serve as an  
example of R-T instability in action as described by Redman et 
al.~(\cite{redman}).

For highly supersonic flows multiple shocks may develop 
(Falle, \cite{falle}),
but at that time the shells are gravitationally stable due to squeezing 
connected to the fast expansion. Later, when they decelerate to 
velocities less than $50\ \mathrm{km\, s}^{-1}$, the gravity starts to be important, 
while
radiative instabilities of the outer shock described by Strickland and Blondin~
(\cite{strick}) loose their influence. 

This work may be extended to nonspherical oscillations using  
the formalism
worked out by Bi\v c\' ak \& Schmidt~(\cite{bicak}) for cosmological 
applications. Here we also ignore the 
deviations from spherical symmetry resulting from initial asymmetry of the 
energy input. In a smooth medium with only large scale density gradients 
the shell
approaches quickly the spherical symmetry as demonstrated by Bisnovatyi-Kogan
and Blinnikov~(\cite{bisnovatyi2}).  
Radial perturbations resulting from
inhomogeneities of the ambient medium and variations of the shell surface 
density will be discussed in a subsequent paper.

\section{The expanding shell in a static, homogeneous medium} 
  
The energy input from an OB association or other sources 
creates a blast-wave propagating into the ambient
medium (Ostriker \& McKee 1988; 
Bisnovatyi-Kogan \& Silich 1995). Since its radius is during the majority 
of the evolution much larger than its thickness, the thin 
shell approximation can be used.
The blast-wave is considered as an expanding infinitesimally thin 
layer surrounding the hot medium inside.
The analytical solution of the expansion in the thin shell approximation 
was derived by Sedov (1959). 
In a static, homogeneous medium without the 
external or internal gravitational field, the blast-wave
is always spherically symmetric, it sweeps the ambient matter
and decelerates. 
Its evolution is given by the equation of motion
\begin{equation}
{\mathrm{d} \over \mathrm{d}t} ( M V ) = S ( P_\mathrm{int} - P_\mathrm{ext} ),
\end{equation}
the equation of mass conservation
\begin{equation}
{\mathrm{d} \over \mathrm{d}t} M = S \rho_0 V,
\end{equation} 
and the equation of state
\begin{equation}
P_\mathrm{int} = {2 \over 3} {E_\mathrm{th} \over V\!ol},
\end{equation}
where $M$ and $V$ are the total mass collected in the shell and
its expansion velocity, $S$ is the shell surface, $P_\mathrm{int}$ and $P_\mathrm{ext}$
are pressures inside and outside of the shell, $\rho_0$ is the density of
the ambient medium and $E_\mathrm{th}$ and $V\!ol$ are the thermal energy and volume
inside of the shell.
Neglecting the external pressure and keeping the $E_\mathrm{th} = \mathrm{const}$, and 
assuming $E_\mathrm{th} = {3 \over 5} E_\mathrm{tot}$, where $E_\mathrm{tot}$ is the total input energy
from the source, 
the radius of the shell $R$ grows with time as 
\begin{eqnarray}
\left({R \over \mathrm{pc}}\right)\ = & \ 72.2 \left({E_\mathrm{tot} \over 10^{51}
\mathrm{erg}}\right)^{1/5} \nonumber\\
 & \left({\rho_0 \over 1 \mathrm{cm}^{-3}}\right)^{-1/5} \left({t \over \mathrm{Myr}}\right)
^{2/5},
\label{s1}
\end{eqnarray}
it decelerates as
\begin{eqnarray}
\left({V \over \mathrm{km\, s}^{-1}}\right)\ = & \ 28.2 
\left({E_\mathrm{tot} \over 10^{51} \mathrm{erg}}\right)^{1/5} \nonumber\\ 
& \left({\rho_0 \over 1 \mathrm{cm}^{-3}}\right)^{-1/5} \left({t \over \mathrm{Myr}}\right)^{-3/5},
\label{s2}
\end{eqnarray}
and its unperturbed surface density $\Sigma_0 $ grows with time as
\begin{eqnarray}
\left({\Sigma_0 \over \mathrm{M_{\odot }}\, \mathrm{pc}^{-2}}\right) = & \ 0.57 
\left({E_\mathrm{tot} \over 10^{51} \mathrm{erg}}\right)^{1/5} \nonumber\\
& \left({\rho_0 \over 1 \mathrm{cm}^{-3}}\right)^{4/5} 
\left({t \over \mathrm{Myr}}\right)^{2/5}. 
\label{s3}
\end{eqnarray} 
 
The linear analysis of the gravitational instability of expanding shells
by Elmegreen (1994) gives for the instantaneous maximum growth rate 
$\omega_\mathrm{BGE}$
of a transversal perturbation of a shell
\begin{equation}
\omega_\mathrm{BGE} = -{3 V \over R} + \left[{V^2 \over R^2} + \left( \pi G \Sigma_0
\over c \right)^2\right]^{1/2},
\label{linstab}
\end{equation}
where $c$ is the sound speed within the shell. The instability occurs for
$\omega_\mathrm{BGE} > 0$.
Inserting Eqs. (\ref{s1} - \ref{s3}) into Eq. (\ref{linstab}), we may derive the time
$t_\mathrm{b}$ when the instability occurs for the first time. 
$\omega _\mathrm{BGE}(t_\mathrm{b}) = 0 $ for 
\begin{eqnarray}
\left({t_\mathrm{b} \over \mathrm{Myr}}\right) = & 28.8 
\left({c \over \mathrm{km\, s}^{-1}}\right)^{5/7} \nonumber\\ 
& \left({E_\mathrm{tot} \over 10^{51} \mathrm{erg}}\right)^{-1/7} 
\left({\rho_0 \over 1 \mathrm{cm}^{-3}}\right)^{-4/7}. 
\end{eqnarray}
The ratio of the wavelength $\lambda $ of the fastest perturbation to $R$
\begin{equation}
\lambda / R = {2 c^2 \over G R \Sigma_0}
\end{equation}
is at $t \ge t_\mathrm{b}$ less than $ {\pi c \over \sqrt{2} V}$. 
The sound speed $c$
in the dense and cold shell is always smaller than the speed of sound in the 
ambient medium, which is smaller than the expansion speed of the shell $V$. 
Therefore, $\lambda / R << 1$.

In a subsequent paper, we shall also discuss 
the instability in non-spherical shells: the values of $R,\ V,$ and
$ \Sigma_0 $ will be  taken from numerical simulations.

\section{Hydrodynamical and Poisson equations on the surface of the shell}

We consider the cold and thin shell of radius $R$ surrounding the 
hot interior 
and expanding with velocity $V$ into a uniform medium of density $\rho_0$.
The intrinsic surface density of the shell $\Sigma$ is composed
of unperturbed part $\Sigma_0$ plus the perturbation $\Sigma_1$
($\Sigma=\Sigma_0 + \Sigma_1$). 
Perturbation $\Sigma_1$ results from the flows on the surface of 
the shell redistributing the accumulated mass. 
We assume that $\Sigma_0$ corresponds to $R$ as $\Sigma_0=
\rho_0 R/3$, which means that all the encountered mass is
accumulated to the shell.
(It comes from $\Sigma_0 = \frac{4} {3}\pi R^3 \rho_0
/ 4\pi R^2$).

\begin{figure} 
\resizebox{\hsize}{!}{\includegraphics{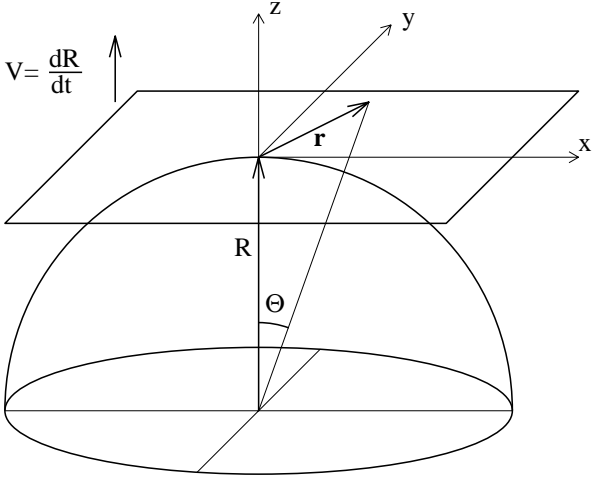}}
\caption{The coordinates on the shell
surface: $\vec\Theta=\vec r/R$
for the position and the angular velocity $\vec\Omega = \vec v/R$
for the surface flows.} 
\label{coord}
\end{figure}

The mass conservation law in a small
area on the surface of the shell is
\begin{equation} \label{mc}
\frac{\partial m}{\partial t}+\left( \nabla, m\vec v \right) = A\  V \rho_0,
\end{equation}
where $m$ is mass in the area $A = 4\pi\alpha R^2$, 
$\alpha$ is a fixed small fraction of the sphere.
$\vec v$ denotes a 
two-dimensional velocity of surface flows connected to perturbations
above the stretching of the area due to expansion of the unperturbed
shell: $\vec r$ and
$\vec v = \dot{\vec r}$ are two-dimensional vectors in the
tangential plane of the shell at the central point of the
considered area. 
We consider
angular coordinates $\vec\Theta=\vec r/R$ and angular
velocity $\vec\Omega = \vec v/R$ to describe the evolution on the surface
of the shell (see Fig.~\ref{coord}). 

With $\Sigma = m / A$ we obtain continuity equation in a form
\begin{equation} \label{kont}
\frac{\partial \Sigma}{\partial t}+2\Sigma\frac{V}{R}
+\Sigma R \left( \nabla, \vec\Omega \right)+R\left(\vec\Omega, \nabla\right) \Sigma 
= V \rho_0 \ ,
\end{equation}

The equation of motion for flows on the shell surface has a form
\begin{equation} 
\frac{1}{A}\frac{\mathrm{d}(m\vec v)}{\mathrm{d}t}=-c^2\nabla\Sigma-\Sigma\nabla
\Phi,
\label{motion}
\end{equation}
where $c$ is the constant isothermal sound speed inside the cold shell, $\Phi$
is the gravitational potential generated by the mass distribution given
in the tangential plane to the shell. 
The shell is confined by the pressure of the bubble interior and 
the outer shock. We assume that the pressure distribution through the shell
corresponds to constant isothermal sound speed $c$ and constant volume density
inside of the shell with a slight deviations close to its inner edge, which
is in contact with the hot interior.
Using the continuity equation 
(\ref{kont}), we can rewrite the Eq. (\ref{motion}) as

\begin{eqnarray} \label{poh}
&&
R\frac{\partial\vec\Omega}{\partial t} + 
V\vec\Omega+3V\vec\Omega\frac{\Sigma_0}{\Sigma}
+ R^2 \left( \vec\Omega, \nabla \right) \vec\Omega 
- R^2 \vec\Omega \left( \nabla, \vec\Omega \right)
\nonumber\\
&&
=-\frac{c^2}{\Sigma}\nabla\Sigma-\nabla\Phi
\end{eqnarray}

The gravitational potential $\Phi$ is related to the surface
density by the Poisson equation
\begin{equation} \label{poisson}
\Delta\Phi = 4\pi G\Sigma\delta(z)\ ,
\end{equation}
where $G$ is the constant of gravitation and $\delta(z)$ is
a delta function of the space coordinate $z$ perpendicular to the surface
of the shell.

\section{Perturbation analysis}

We assume a small perturbation of the shell surface density
$\Sigma_1 \ll \Sigma_0$ which evolves due to surface 
flows given with velocity $\vec v$. 
The perturbed hydrodynamical equations (\ref{kont}), (\ref{poh}) and
the perturbed Poisson equation (\ref{poisson}) have form

\begin{equation} \label{pkont}
{\partial \Sigma_1 \over \partial t} + 2\Sigma_1 {V \over R} 
+ \Sigma R \left( \nabla, \vec \Omega \right) + R \left( \vec \Omega,
\nabla \right) \Sigma_1 = 0, 
\end{equation}

\begin{eqnarray} \label{ppoh}
&&
R {\partial \vec \Omega \over \partial t}
+ R^2 \left( \vec\Omega, \nabla \right) \vec\Omega 
- R^2 \vec\Omega \left( \nabla, \vec\Omega \right)
\nonumber\\
&&
= -{c^2 \over \Sigma_0}(1-{\Sigma_1 \over \Sigma_0}) \nabla
\Sigma_1
- \nabla \Phi_1 - 4V\vec \Omega + 3V \vec \Omega {\Sigma_1 
\over \Sigma_0},
\end{eqnarray}

\begin{equation}
\Delta \Phi_1 = 4\pi G \Sigma_1 \delta (z),
\label{poisson1}
\end{equation}

\noindent
where the $1/\Sigma$ in Eq. (\ref{poh}) was evaluated up 
to quadratic terms in $\Sigma_1$
: ${c^2 \over \Sigma}\nabla \Sigma =
{c^2 \over \Sigma_0} (1 - {\Sigma_1 \over \Sigma_0} +$ higher order terms
$)\ \nabla \Sigma_1$, and $\nabla \Sigma_0 = 0$ on the shell surface.

The perturbation of the surface density $\Sigma_1$ and the angular velocity
of the surface flows $\vec\Omega$ can be written as
\begin{eqnarray}\label{fourier_t}
\Sigma_1=\Sigma_{10} + \sum_{\vec\eta} \Sigma_{\vec\eta} \mathrm{e}^{i \left(\vec{\eta},
\vec{\Theta} \right) } \ ,\nonumber\\
\vec\Omega=\vec\Omega_0+\sum_{\vec\eta} \vec\Omega_{\vec\eta} \mathrm{e}^{i
\left( \vec{\eta}, \vec{\Theta}\right)} \ ,
\end{eqnarray}
which may be inserted to the perturbed Eqs. (\ref{pkont}) and 
(\ref{ppoh}).
$\vec\eta$ denotes a dimensionless wave-vector $\vec\eta=\vec kR$.
We assume no surface macroscopic flow through all the considered area 
which means $\vec\Omega_0=0$. Further we assume $\Sigma_{10}=0$ (mass
accumulation due to expansion to the ambient medium is included in  
$\Sigma_0$).
The Fourier transform of the Eq. (\ref{pkont}) is

\begin{eqnarray}\label{fkont}
\dot\Sigma_{\vec\eta} 
 & + \Sigma_0\left(-i \vec\eta, \vec\Omega_{-\vec\eta}\right)
+ 2\frac{V}{R} \Sigma_{\vec\eta} 
 + \sum_{\vec\eta'} \left(-i\vec\eta', \vec\Omega_
  {-\vec\eta'}\right) \Sigma_{\vec\eta - \vec\eta'} \nonumber\\
& + \sum_{\vec\eta'}\left(\vec\Omega_{\vec\eta-\vec\eta'},
   -i\vec\eta'\right) \Sigma_{\vec\eta'}=0 
\end{eqnarray}
where we used the identity $\vec\eta'+(\vec\eta-\vec\eta')=\vec\eta$.

The solution of the Poisson equation (\ref{poisson1}) 
\begin{equation}
\nabla\Phi_{1} = -2\pi G\sum_{\vec\eta}\Sigma_{\vec\eta}
\frac{i\vec\eta}{|\vec\eta|} \mathrm{e}^{i\left(\vec\eta, \vec\Theta\right)}
\end{equation}
can be inserted to the Fourier transform of the Eq. (\ref{ppoh}).
We get:

\begin{eqnarray}\label{fpoh}
& &
R\dot{\vec\Omega}_{\vec\eta} 
+ R\sum_{\vec\eta'} \left( \vec\Omega_{\vec\eta-\vec\eta'}, -i\vec\eta' \right)
\vec\Omega_{\vec\eta'}
- R\sum_{\vec\eta'} \vec\Omega_{\vec\eta-\vec\eta'} 
\nonumber\\
& &
\left(-i\vec\eta', \vec\Omega_{-\vec\eta'} \right)
= - \frac{c^2}{R \Sigma_0}i\vec\eta \Sigma_{\vec\eta}
+ \frac{c^2}{R \Sigma_0^2}\sum_{\vec\eta'}i\vec\eta'
\Sigma_{\vec\eta-\vec\eta'}\Sigma_{\vec\eta'} 
\nonumber\\ 
& &
+ 2\pi G\frac{i\vec\eta}{|\vec\eta|}\Sigma_{\vec\eta} 
- 4V\vec\Omega_{\vec\eta}
+\frac{3V}{\Sigma_0}\sum_{\vec\eta'}\vec\Omega_{\vec\eta'}
\Sigma_{\vec\eta-\vec\eta'}.
\end{eqnarray}

\section{Linear analysis}
We give the solution of Eqs. (\ref{fkont}) and (\ref{fpoh}) using the
linear terms only. This linear solution will be later compared to the results 
obtained with the nonlinear terms.
The complete linear analysis of the expanding shell was also done by
Elmegreen~(\cite{elmegreen}).
  
Linearized Eqs. (\ref{fkont}) and (\ref{fpoh}) have a form
\begin{equation}
\dot\Sigma_{\vec\eta} 
+ \Sigma_0 \left( -i\vec\eta, \vec\Omega_{-\vec\eta}\right) 
+ 2\frac{V}{R} \Sigma_{\vec\eta} = 0
\end{equation}

\begin{equation}
R\dot{\vec\Omega}_{\vec\eta} =
- \frac{c^2}{\Sigma_0}i\vec k \Sigma_{\vec\eta}
+ 2\pi G\frac{i\vec\eta}{|\vec\eta|}\Sigma_{\vec\eta} 
- 4V\vec\Omega_{\vec\eta}.
\end{equation}

Angular velocity $\vec\Omega_{\vec\eta}$ can be split in two
components parallel and orthogonal to the wave-vector $\vec\eta$
\begin{equation}\label{veldis}
\vec\Omega_{\vec\eta}=\Omega_{\vec\eta_\parallel}
\frac{\vec\eta}{|\vec\eta|}
+\Omega_{\vec\eta_\perp}\frac{\vec\eta_\perp}{|\vec\eta_\perp|} \ ,
\end{equation}
where $\vec\eta_\perp$ is a vector in the tangential plane, 
which is perpendicular to the wave-vector $\vec\eta$
and $|\vec\eta_{\perp}|=|\vec\eta|$.

We get the set of equations

\begin{equation}\label{linkont}
\dot{\Sigma}_{\vec\eta}=-2\frac{V}{R}\Sigma_{\vec\eta}
-i\eta\Sigma_0\Omega_{\vec\eta_\parallel}
\end{equation}

\begin{equation}\label{linpohpar}
\dot{\Omega}_{\vec\eta_\parallel}=(i\frac{2\pi G}{R}
-i\frac{c^2\eta}{\Sigma_0R^2})\Sigma_{\vec\eta}
-4\frac{V}{R}\Omega_{\vec\eta_\parallel}
\end{equation}

\begin{equation}\label{linpohperp}
\dot{\Omega}_{\vec\eta_\perp}=-4\frac{V}{R}\Omega_{\vec\eta_\perp}
\end{equation}
which can be formally written as
\begin{equation}\label{linform}
\pmatrix{
\dot\Sigma_{\vec\eta} \cr 
\dot\Omega_{\vec\eta_\parallel} \cr
\dot\Omega_{\vec\eta_\perp}
} = 
\cal L \rm
\pmatrix{
\Sigma_{\vec\eta} \cr 
\Omega_{\vec\eta_\parallel} \cr
\Omega_{\vec\eta_\perp}
}
\ .
\end{equation}

We find the eigenvalues and eigenvectors of the linear operator $\cal L$.
\begin{equation}\label{omega12}
\omega_{\eta}^{(1,2)} = i\frac{3V}{R} \pm \sqrt{-\frac{V^2}{R^2} +
\frac{\eta^2 c^2}{R^2} - \frac{2\pi G\Sigma_0\eta}{R}}
\end{equation}

\begin{equation}
\omega_{\eta}^{(3)}=i4\frac{V}{R}  \ ,
\end{equation}
where $\eta $ is the magnitude of $\vec\eta $.

The related eigenvectors are

\begin{equation}\label{eigenvect}
\pmatrix{\Sigma_{\vec\eta} \cr \Omega_{\vec\eta_\parallel} \cr
\Omega_{\vec\eta_\perp}}^{(1,2)} =  \pmatrix{\eta\Sigma_0 \cr
i 2\frac{V}{R} - \omega^{(1,2)}_{\eta } \cr 0} 
\end{equation}

\begin{equation}
\pmatrix{\Sigma_{\vec\eta} \cr \Omega_{\vec\eta_\parallel} \cr
\Omega_{\vec\eta_\perp}}^{(3)} = \pmatrix{0\cr 0\cr 1} \ .
\end{equation}

\begin{figure}
\resizebox{\hsize}{!}{\includegraphics{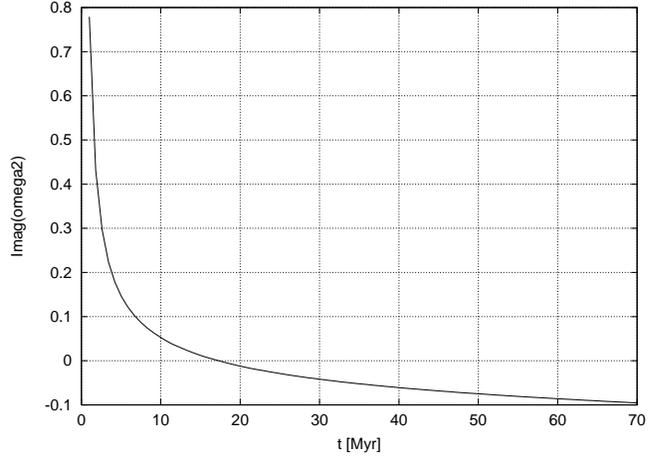}}
\caption{The time dependence of the imaginary part of the $\omega^{(2)}$,
which can cause the instability. The Sedov solution was used with following 
parameters: total energy $E_{tot} = 10^{53} \mathrm{erg}$, density of ambient medium 
$n_0 = 1~\mathrm{cm}^{-3}$, average molecular weight $\mu = 1.3$, sound speed in the
shell $c = 1~\mathrm{km\, s}^{-1}$.} 
\label{omegadep}
\end{figure}

The eigenvalues $\omega_{\eta}$ are time dependent and  can be used 
to obtain a criteria
for the instability of the shell. For a short time for which the change 
of the linear
operator $\cal L$ is small (and we can always find the time interval which 
is short enough), 
the solution of Eqs. (\ref{linform}) is a part of the exponential 
function $\sim
\mathrm{e}^{i\omega t}$. The $\omega^{(3)}_{\eta}$ has always meaning of the 
decrease of
perturbations. If $\omega^{(1,2)}_{\eta}$ have got a real part, 
solution is stable with
decreasing oscillations. If not, $\omega^{(1)}_{\eta}$ indicates decrease,
$\omega^{(2)}_{\eta}$ can be imaginary negative and it have meaning 
of the perturbations
growth rate. The time evolution of the imaginary part of the 
$\omega^{(2)}_{\eta}$ is 
shown by Fig.~\ref{omegadep}.

Since the eigenvalues $\omega_{\eta}^{(1,2)}$ depend  on 
$\eta $ through the relation (\ref{omega12}), the maximum 
perturbation growth rate $\omega_{\eta, \mathrm{max}}^{(1,2)}$ can be found:
\begin{equation}
\omega_{\eta,\mathrm{max}}^{(1,2)} = i\frac{3V}{R} \pm \sqrt{-\frac{V^2}{R^2} -
\frac{\pi^2 G^2 \Sigma_0^2 }{c^2}}
\end{equation}
and occurs at the wavenumber:
\begin{equation}
\eta_\mathrm{max}=\frac{\pi G \Sigma_0 R}{c^2} \ .
\end{equation}

$\omega^{(2)}_{\eta, \mathrm{max}}$ is close to $\omega_\mathrm{BGE}$ derived by 
Elmegreen~(\cite{elmegreen}), 
actually $\omega^{(2)}_{\eta, \mathrm{max}} = - i \omega_\mathrm{BGE}$.
If the shell is unstable, i. e. the imaginary part of the
$\omega^{(2)}_{\eta }$ is negative, for the certain time, the shell may
break to fragments. In Ehlerov\'a et al. (1997)
is the fragmentation integral for the maximum perturbation growth rate 
defined as:
\begin{equation}
I_\mathrm{f}(t) \equiv \int_{t_\mathrm{b}}^{t} - i\omega^{(2)}_{\eta,\mathrm{max}}(t')
\mathrm{d}t' \ ,
\end{equation}
where $t_\mathrm{b}$ is the time when the instability begins. 
We generalize it for any unstable mode with the wavenumber $\eta $ using an 
analogy to the value for $\omega_\mathrm{BGE}$ as
\begin{equation}
I_{\mathrm{f}, \eta }(t) \equiv \int_{t_\mathrm{b}}^{t} - i\omega^{(2)}_{\eta }(t')
\mathrm{d}t' \ ,
\label{genfrag}
\end{equation} 
The fragmentation 
time $t_{\mathrm{f}, \eta }$
(the time giving some development level of a fragment) is defined as 
the time when
the fragmentation integral is equal to one:
\begin{equation}
I_{\mathrm{f}, \eta }(t_{\mathrm{f}, \eta }) = \int_{t_\mathrm{b}}^{t_{\mathrm{f}, \eta }} 
- i\omega^{(2)}_{\eta } (t') \mathrm{d}t' = 1\ .
\end{equation}

\section{Nonlinear analysis}
The equations with nonlinear terms are solved by the similar procedure 
as adopted by
Fuchs~(\cite{fuchs}).
We rewrite the nonlinear Eqs. (\ref{fkont}) and (\ref{fpoh})
in the form

\begin{equation}\label{nonlin}
\pmatrix{
\dot\Sigma_{\vec\eta} \cr 
\dot\Omega_{\vec\eta_\parallel} \cr
\dot\Omega_{\vec\eta_\perp}
} = 
\cal L \rm
\pmatrix{
\Sigma_{\vec\eta} \cr 
\Omega_{\vec\eta_\parallel} \cr
\Omega_{\vec\eta_\perp}
}
+ {\cal N} \ ,
\end{equation}
where ${\cal L}$ is the linear part and ${\cal N}$ represents
the non-linear terms. We search for a solution of Eqs. (\ref{nonlin})
as a combination of the eigenvectors obtained from the 
previous linear analysis

\begin{eqnarray}\label{ansatz}
\pmatrix{
\Sigma_{\vec\eta} \cr 
\Omega_{\vec\eta_\parallel} \cr 
\Omega_{\vec\eta_\perp}}
& = & 
\psi_{\vec\eta}(t) \pmatrix{\Sigma_{\vec\eta} \cr 
\Omega_{\vec\eta_\parallel} \cr
\Omega_{\vec\eta_\perp}}^{(1)}
+
\nonumber\\
& & +\xi_{\vec\eta}(t) \pmatrix{\Sigma_{\vec\eta} \cr 
\Omega_{\vec\eta_\parallel} \cr
\Omega_{\vec\eta_\perp}}^{(2)}
+\phi_{\vec\eta}(t) \pmatrix{\Sigma_{\vec\eta} \cr 
\Omega_{\vec\eta_\parallel} \cr
\Omega_{\vec\eta_\perp}}^{(3)},
\end{eqnarray}
where $\psi_{\vec\eta(t)}$, $\xi_{\vec\eta(t)}$ and 
$\phi_{\vec\eta(t)}$ are time dependent amplitudes
of eigenvectors. 
There are always four solutions differing in eigenvectors multiplied by 
$\pm i$ or $\pm 1$. We select only solutions with physical relevance given by
eigenvectors (\ref{eigenvect}).
We find orthonormal vectors 
$(\Sigma_{\vec\eta},\Omega_{\vec\eta_\parallel},
\Omega_{\vec\eta_\perp})^{(1,2,3)}$
in order that

\begin{eqnarray}
& & (\Sigma_{\vec\eta},
\Omega_{\vec\eta_\parallel},\Omega_{\vec\eta_\perp})^{(j)}
\pmatrix{
\Sigma_{\vec\eta} \cr 
\Omega_{\vec\eta_\parallel} \cr
\Omega_{\vec\eta_\perp}
}^{(j)}=1
\nonumber\\
& & (\Sigma_{\vec\eta},
\Omega_{\vec\eta_\parallel},\Omega_{\vec\eta_\perp})^{(j)}
\pmatrix{
\Sigma_{\vec\eta} \cr 
\Omega_{\vec\eta_\parallel} \cr
\Omega_{\vec\eta_\perp}
}^{(k\not=j)}=0,
\end{eqnarray}
where $j, k = 1,2,3$. The orthonormal vectors are

\begin{eqnarray} \label{orthonorm}
\begin{array}{l}
(\Sigma_{\vec\eta},\Omega_{\vec\eta_\parallel},\Omega_{\vec\eta_\perp})^{(1)}=
\left(
\frac{\omega^{(2)}-i2\frac{V}{R}}{\eta\Sigma_0(\omega^{(2)}-\omega^{(1)})}
,\frac{1}{(\omega^{(2)}-\omega^{(1)})} 
,0
\right)
\\
(\Sigma_{\vec\eta},\Omega_{\vec\eta_\parallel},\Omega_{\vec\eta_\perp})^{(2)}=
\left(
\frac{\omega^{(1)}-i2\frac{V}{R}}{\eta\Sigma_0(\omega^{(1)}-\omega^{(2)})}
,\frac{1}{(\omega^{(1)}-\omega^{(2)})} 
,0
\right)
\\
(\Sigma_{\vec\eta},\Omega_{\vec\eta_\parallel},\Omega_{\vec\eta_\perp})^{(3)}=
\left(
0,0,1
\right)
\end{array}
\ .
\end{eqnarray}

We insert ansatz (\ref{ansatz}) into Eq. (\ref{nonlin}),
multiply it by the orthonormal vectors  (\ref{orthonorm})
and obtain a set of equations for amplitudes $\psi_{\vec\eta(t)}$,
$\xi_{\vec\eta(t)}$ and  $\phi_{\vec\eta(t)}$

\begin{eqnarray}\label{psi}
\dot\psi_{\vec\eta} & = & i\omega^{(1)}\psi_{\vec\eta} 
- \left(
\frac{\partial}{\partial t}{\pmatrix{
\Sigma_{\vec\eta} \cr 
\Omega_{\vec\eta_\parallel} \cr
\Omega_{\vec\eta_\perp}
}}^{(1)},
\left(\Sigma_{\vec\eta},\Omega_{\vec\eta_\parallel},\Omega_{\vec\eta_\perp}\right)^{(1)}
\right)
\psi_{\vec\eta}
\nonumber\\
& & 
- \left(
\frac{\partial}{\partial t}{\pmatrix{
\Sigma_{\vec\eta} \cr 
\Omega_{\vec\eta_\parallel} \cr
\Omega_{\vec\eta_\perp}
}}^{(2)}, 
\left(\Sigma_{\vec\eta},\Omega_{\vec\eta_\parallel},\Omega_{\vec\eta_\perp}\right)^{(1)}
\right)
\xi_{\vec\eta}
\\
& & + \left(\cal N \rm,
\left(\Sigma_{\vec\eta},\Omega_{\vec\eta_\parallel},\Omega_{\vec\eta_\perp}\right)^{(1)}
\right)
\nonumber
\end{eqnarray}

\begin{eqnarray} \label{xi}
\dot\xi_{\vec\eta} & = & i\omega^{(2)}\xi_{\vec\eta} 
- \left(
\frac{\partial}{\partial t}{\pmatrix{
\Sigma_{\vec\eta} \cr 
\Omega_{\vec\eta_\parallel} \cr
\Omega_{\vec\eta_\perp}
}}^{(2)}, 
\left(\Sigma_{\vec\eta},\Omega_{\vec\eta_\parallel},\Omega_{\vec\eta_\perp}\right)^{(2)}
\right)
\xi_{\vec\eta}
\nonumber\\
& & 
- \left(
\frac{\partial}{\partial t}{\pmatrix{
\Sigma_{\vec\eta} \cr 
\Omega_{\vec\eta_\parallel} \cr
\Omega_{\vec\eta_\perp}
}}^{(1)}, 
\left(\Sigma_{\vec\eta},\Omega_{\vec\eta_\parallel},\Omega_{\vec\eta_\perp}\right)^{(2)}
\right)
\psi_{\vec\eta}
\\
& & + \left(\cal N \rm,
\left(\Sigma_{\vec\eta},\Omega_{\vec\eta_\parallel},\Omega_{\vec\eta_\perp}\right)^{(2)}
\right)
\nonumber
\end{eqnarray}

\begin{eqnarray}\label{phi}
\dot\phi_{\vec\eta} & = & i\omega^{(3)}\phi_{\vec\eta} 
- \left(
\frac{\partial}{\partial t}{\pmatrix{
\Sigma_{\vec\eta} \cr 
\Omega_{\vec\eta_\parallel} \cr
\Omega_{\vec\eta_\perp}
}}^{(3)}, 
\left(\Sigma_{\vec\eta},\Omega_{\vec\eta_\parallel},\Omega_{\vec\eta_\perp}\right)^{(3)}
\right)
\phi_{\vec\eta}
\nonumber\\
& & + \left(\cal N \rm,
\left(\Sigma_{\vec\eta},\Omega_{\vec\eta_\parallel},\Omega_{\vec\eta_\perp}\right)^{(3)}
\right),
\end{eqnarray}
where we denote $\omega^{(j)} \equiv \omega^{(j)}_{\eta, \mathrm{max}}$. 

The Eq. (\ref{phi}) is decoupled from the others and 
its solution is the decrease of the
initial value of $\phi$.
Eqs. (\ref{psi}) and (\ref{xi}) are coupled through the 
linear and nonlinear terms. The interaction through 
the linear terms is weak, 
since the  coupled linear terms have smaller amplitudes compared to
linear terms and they decrease with time due to their dependence on the time
derivatives of the eigenvectors, which are very small in the later stages of 
the shell evolution. The coupling through the nonlinear terms leads to the
terms of the third and higher orders, which can be neglected with respect to
quadratic terms.
Furthermore, the solution of the Eq. (\ref{psi}) has
a decreasing character, because the first term on the right side, which 
includes the ``stable'' $\omega^{(1)}$, dominates.
The Eq. (\ref{xi}) is the most interesting one, because it has $\omega^{(2)}$ in
the first linear term, and only the $\omega^{(2)}$ can be imaginary negative,
which has meaning of instability.
The explicit form of the Eg. (\ref{xi}) is

\begin{eqnarray}\label{bigequ}
\dot\xi_{\vec\eta} & = & i\omega^{(2)}\xi_{\vec\eta}
-\frac{(|\dot{\vec\eta}|\Sigma_0+|\vec\eta|\dot\Sigma_0)(\omega^{(1)} - i 2\frac{V}{R})
+|\vec\eta|\Sigma_0}{|\vec\eta|\Sigma_0(\omega^{(1)}-\omega^{(2)})}
\nonumber\\
& & 
\frac{(i 2\frac{\dot VR-V^2}{R^2} - \dot{\omega}^{(2)})}{\dots}
\xi_{\vec\eta}
-\frac{\omega^{(1)}-i2\frac{V}{R}}
{|\vec\eta|\Sigma_0\left( \omega^{(1)}-\omega^{(2)}\right) }
\sum_{\vec\eta'}
\nonumber\\
& &
i|\vec\eta-\vec\eta'| \Sigma_0 
\left( \psi_{\vec\eta-\vec\eta'}+\xi_{\vec\eta-\vec\eta'}\right)
\left[
\left( i2\frac{V}{R}-\omega^{(1)}\right) \psi_{\vec\eta'}
\right.
\nonumber\\
& &
\left.
+\left( i2\frac{V}{R}-\omega^{(2)}\right) \xi_{\vec\eta'}
\right]
|\vec\eta'|
-\frac{\omega^{(1)}-i2\frac{V}{R}}
{|\vec\eta|\Sigma_0\left( \omega^{(1)}-\omega^{(2)}\right) }
\nonumber\\
& &
\sum_{\vec\eta'} i |\vec\eta'| \Sigma_0
\left( \psi_{\vec\eta'}+\xi_{\vec\eta'}\right)
\left\{
\left[
\left( i2\frac{V}{R}-\omega^{(1)}\right) \psi_{\vec\eta-\vec\eta'}
\right.
\right.
\nonumber\\
& &
\left.
\left.
+\left( i2\frac{V}{R}-\omega^{(2)}\right) \xi_{\vec\eta-\vec\eta'}
\right]
\frac{(\vec\eta-\vec\eta' , \vec\eta')}{|\vec\eta-\vec\eta'|}
+
\right.
\\
& &
\left.
\phi_{\vec\eta-\vec\eta'}
\frac{(\vec\eta-\vec\eta'_{\perp} ,
\vec\eta')}{|\vec\eta-\vec\eta'_{\perp}|}
\right\}
+\frac{3V}{R\Sigma_0 \left( \omega^{(1)}-\omega^{(2)}\right)}
\sum_{\vec\eta'}
\nonumber\\
& &
|\vec\eta-\vec\eta'| \Sigma_0 
\left( \psi_{\vec\eta-\vec\eta'}+\xi_{\vec\eta-\vec\eta'}\right)
\left\{
\left[
\left( i2\frac{V}{R}-\omega^{(1)}\right) \psi_{\vec\eta'}
\right.
\right.
\nonumber\\
& &
\left.
\left.
+\left( i2\frac{V}{R}-\omega^{(2)}\right) \xi_{\vec\eta'}
\right]
\frac{(\vec\eta' , \vec\eta)}{|\vec\eta'||\vec\eta|}
+\phi_{\vec\eta'}
\frac{(\vec\eta'_{\perp} , \vec\eta)}{|\vec\eta'_{\perp}||\vec\eta|}
\right\}
+
\nonumber\\
& &
- \frac{i}{\left( \omega^{(1)}-\omega^{(2)}\right)}
\sum_{\vec\eta'}
\left\{ \left[ 
\left( i2\frac{V}{R}-\omega^{(1)}\right) \psi_{\vec\eta-\vec\eta'} +
\right.
\right.
\nonumber\\
& &
\left.
\left.
\left( i2\frac{V}{R}-\omega^{(2)}\right) \xi_{\vec\eta-\vec\eta'}
\right]
\frac{(\vec\eta-\vec\eta', \vec\eta')}{|\vec\eta-\vec\eta'|}
+ \phi_{\vec\eta-\vec\eta'} 
\right.
\nonumber\\
& &
\left.
\frac{(\vec\eta-\vec\eta'_{\perp}, \vec\eta')}{|\vec\eta-\vec\eta'_{\perp}|}
\right\}
\left\{ \left[ 
\left( i2\frac{V}{R}-\omega^{(1)}\right) \psi_{\vec\eta'} +
\left( i2\frac{V}{R}-
\right.
\right.
\right.
\nonumber\\
& &
\left.
\left.
\left.
\omega^{(2)}\right) \xi_{\vec\eta'} \right]
\frac{(\vec\eta', \vec\eta)}{|\vec\eta'||\vec\eta|}
+ \phi_{\vec\eta'}
\frac{(\vec\eta'_{\perp}, \vec\eta)}{|\vec\eta'_{\perp}||\vec\eta|}
\right\}
- \frac{i}{\left( \omega^{(1)}-\omega^{(2)}\right)}
\nonumber\\
& &
\sum_{\vec\eta'}
\left\{ \left[ 
\left( i2\frac{V}{R}-\omega^{(1)}\right) \psi_{\vec\eta-\vec\eta'} +
\left( i2\frac{V}{R}-\omega^{(2)}\right) 
\right.
\right.
\nonumber\\
& &
\left.
\left.
\xi_{\vec\eta-\vec\eta'} \right]
\frac{(\vec\eta-\vec\eta', \vec\eta)}{|\vec\eta-\vec\eta'||\vec\eta|}
+ \phi_{\vec\eta-\vec\eta'} 
\frac{(\vec\eta-\vec\eta'_{\perp}, \vec\eta)}{|\vec\eta-\vec\eta'_{\perp}||\vec\eta|}
\right\}
\nonumber\\
& &
\left\{ \left[ 
\left( i2\frac{V}{R}-\omega^{(1)}\right) \psi_{\vec\eta'} +
\left( i2\frac{V}{R}-\omega^{(2)}\right) \xi_{\vec\eta'} \right]
|\vec\eta'|
\right.
\nonumber\\
& &
\left.
+ \phi_{\vec\eta'}
|\vec\eta'|
\right\}
+
\frac{i c^2}{R^2 \Sigma_0^2 \left( \omega^{(1)}-\omega^{(2)}\right)}
\sum_{\vec\eta'}
|\vec\eta-\vec\eta'|\Sigma_0
\nonumber\\
& &
\left( \psi_{\vec\eta-\vec\eta'}+\xi_{\vec\eta-\vec\eta'}\right)
|\vec\eta'| \Sigma_0
\left( \psi_{\vec\eta'}+\xi_{\vec\eta'}\right)
\frac{(\vec\eta' , \vec\eta)}{|\vec\eta'|}
\nonumber
\end{eqnarray}

In analogy to Fuchs~(\cite{fuchs}) we group together 
components with fully imaginary $\omega_{\vec\eta}$ into
wave-packets and take the wavenumber $\eta_\mathrm{max}$.
Other components are grouped to the wave-packets of the same width
and approximated by the average wave-numbers.
The $\xi_{\vec\eta_\mathrm{max}}$ modes grow,
the $\psi_{\vec\eta_\mathrm{max}}$ and the $\phi_{\vec\eta_\mathrm{max}}$ modes
descend. Other modes oscillate with decrease.

\begin{figure}
\resizebox{\hsize}{!}{\includegraphics{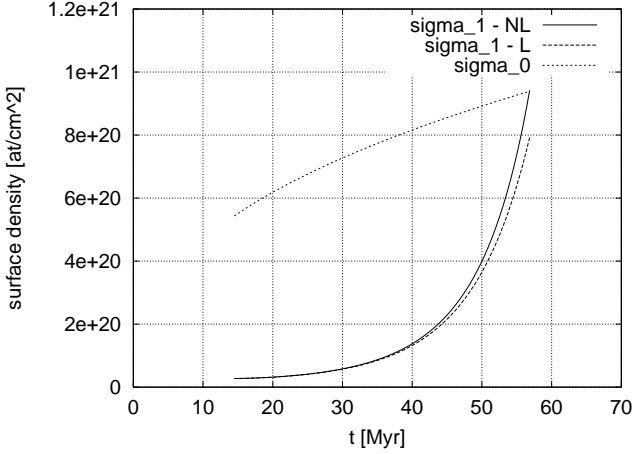}}
\caption{
The evolution of the maximum perturbation of the surface
density in the case when the initial values of linear and nonlinear 
terms of perturbation are in phase.} 
\label{dens1}
\end{figure}

Demanding $\eta=\eta'=\left|\vec\eta-\vec\eta'\right|=\eta_\mathrm{max}$
we obtain from geometrical consideration:

\begin{equation}\label{geometry}
\begin{array}{ll}
 (\vec\eta-\vec\eta'_{\pm}, \vec\eta'_{\pm})=-\frac{1}{2}\eta^2_\mathrm{max}
& 
 (\vec\eta'_{\pm}, \vec\eta)=\frac{1}{2}\eta^2_\mathrm{max}
\end{array}
\ ,
\end{equation}

\noindent
where $\vec\eta'_{+}$ and $\vec\eta'_{-}$ are two wave-vectors inclined at
angles $60^o$ to the wave-vector $\vec\eta$. It means that 
$\xi_{\vec\eta}$ wave-packets non-linearly interact with others
with wave-vectors $\vec\eta'_{\pm}$.
Using Eqs. (\ref{omega12}) and (\ref{geometry}),
the set of Eqs. (\ref{bigequ}) can be simplified to the form

\begin{equation}\label{finalequ}
\begin{array}{lclcccr}
\dot\xi & = & (i\omega^{(2)} + A) \xi & + & B \xi_{+}\xi_{-}
\\
\dot\xi_{+} & = & (i\omega^{(2)} + A) \xi_{+} & + & B \xi\xi_{-}^{*}
\nonumber\\
\dot\xi_{-} & = & (i\omega^{(2)} + A) \xi_{-} & + & B \xi\xi_{+}^{*}
\ ,
\end{array}
\end{equation}
where $\xi \equiv \xi_{\vec\eta}$, $\xi_{+} \equiv \xi_{\vec\eta_{+}}$, 
$\xi_{-} \equiv \xi_{\vec\eta_{-}}$, the asterisk has meaning of the
complex conjunction
and
\begin{eqnarray}
A &=&-\frac{-i18RVc^2\dot\Sigma_0\gamma + 18V^2c^2\dot\Sigma_0
+2R^4\pi^2G^2\rho_0^2\dot\Sigma_0}
{2\Sigma_0(9V^2c^2+\pi^2G^2\rho_0^2R^4)}-
\nonumber\\
& &
-\frac{-i18V^2c^2\Sigma_0\gamma + 2\pi^2G^2\rho_0^2R^3V\Sigma_0}
{2\Sigma_0(9V^2c^2+\pi^2G^2\rho_0^2R^4)}-
\nonumber\\
& &
-\frac{i9R\dot Vc^2\Sigma_0\gamma + 9V\dot Vc^2\Sigma_0}
{2\Sigma_0(9V^2c^2+\pi^2G^2\rho_0^2R^4)}
\end{eqnarray}
and
\begin{equation}
B=\frac{i \pi^3 G^3 \Sigma_0^3 R 
- i 12 \pi G \Sigma_0 c^2 V \left( \frac{V}{R} + i\gamma\right)}
{4c^4\gamma}\ , 
\end{equation}
where
\begin{equation}
\gamma=\sqrt{-\frac{V^2}{R^2}-\frac{\pi^2G^2\Sigma_0^2}{c^2}} \ .
\end{equation}

The set of Eqs. (\ref{finalequ}) describe the time evolution of one
triplet 
of the most interacting modes $(\vec\eta,\vec\eta_{+},\vec\eta_{-})$.
A is coming from the time derivative of the eigenvector in (\ref{xi}). 
Its amplitude is much less than the amplitude of $\omega^{(2)}$ except
around the time $t_\mathrm{b}$, when $\omega^{(2)}$ is close to zero. The resulting 
effect is the time evolution of $\xi, \xi_+, \xi_-$, which for $t > t_\mathrm{b}$ 
initially decrease and start to grow only with some delay (see Fig. 4).
Nevertheless, the transformation of $\xi, \xi_+, \xi_-$ to $\Sigma $ gives
the continuous increase of $\Sigma $ after $t_\mathrm{b}$ (see Fig. 3). A is the term 
originating in the transformation of $\Sigma $, $\Omega $ to 
$\xi, \xi_+, \xi_-$, and it is important around $t_\mathrm{b}$ only.

\begin{figure}
\resizebox{\hsize}{!}{\includegraphics{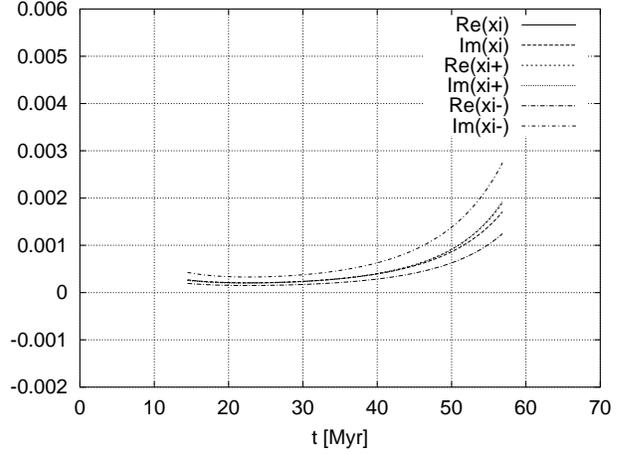}}
\caption{The solution of the set of Eqs. (\ref{finalequ})
with the initial conditions corresponding to Fig.~\ref{dens1}.} 
\label{ampl1}
\end{figure}

\subsection{The numerical solution}

The set of Eqs. (\ref{finalequ}) can be solved numerically.
We start at the time $t_\mathrm{b}$ which is the time when the instability begins 
(imaginary part of $\omega_{\eta,\mathrm{max}}^{(2)}$ starts to be negative).
First we select real and imaginary parts of all 
initial perturbation amplitudes $\xi ,\ \xi _+,\ \xi _-$, 
which have the  meaning of initial perturbations of the surface density
and of the velocity, such that they correspond to 
$\Sigma_1 / \Sigma_0 = 0.05$. 
The magnitude of these perturbations in physical values
can be computed from the eigenvectors (\ref{eigenvect}).

The solution is determined by parameters of two types: the first ones,
as speed of sound $c$ in the shell, are constant values, the second ones, as 
the radius of the shell $R(t)$, the expansion velocity $V(t)$ 
and its time derivative
and the surface density $\Sigma_0(t)$ and its time derivative, are functions 
of time. We can get them either from the analytical Sedov solution (\ref{s1} -
\ref{s3}), 
or from the numerical simulations of the expanding HI shells described by 
Ehlerov\'a et al. (\cite{ehlerova}). In this paper we use the Sedov solution
(Eqs. \ref{s1}-\ref{s3}) with following 
parameters: total energy $E_\mathrm{tot} = 10^{53} \mathrm{erg}$, density of ambient medium 
$n_0 = 1~\mathrm{cm}^{-3}$, average molecular weight $\mu = 1.3$, sound speed in the
shell $c = 1~\mathrm{km\, s}^{-1}$.

\begin{figure}
\resizebox{\hsize}{!}{\includegraphics{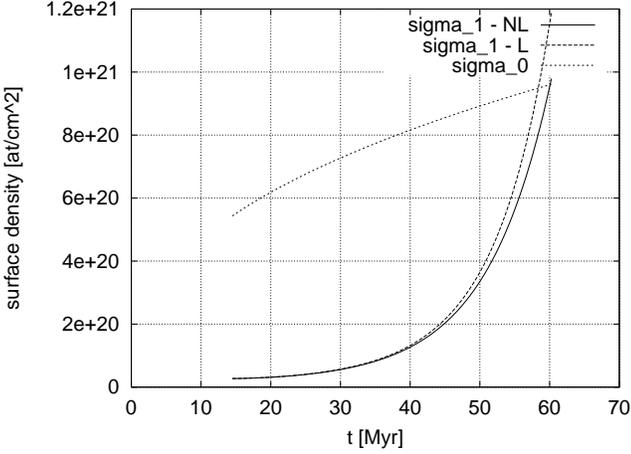}}
\caption{The evolution of the maximum perturbation of the surface
density in the case when the initial values of linear and nonlinear 
terms of perturbation are in anti-phase.} 
\label{dens2}
\end{figure}

\begin{figure}
\resizebox{\hsize}{!}{\includegraphics{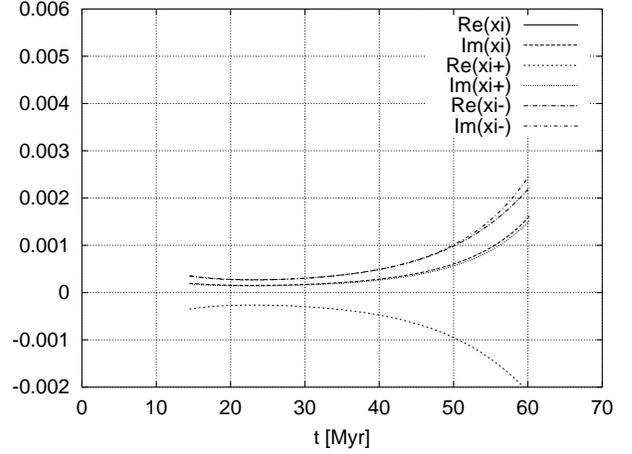}}
\caption{The solution of the set of Eqs. (\ref{finalequ})
with the initial conditions corresponding to Fig.~\ref{dens2}.} 
\label{ampl2}
\end{figure}

\begin{figure}
\resizebox{\hsize}{!}{\includegraphics{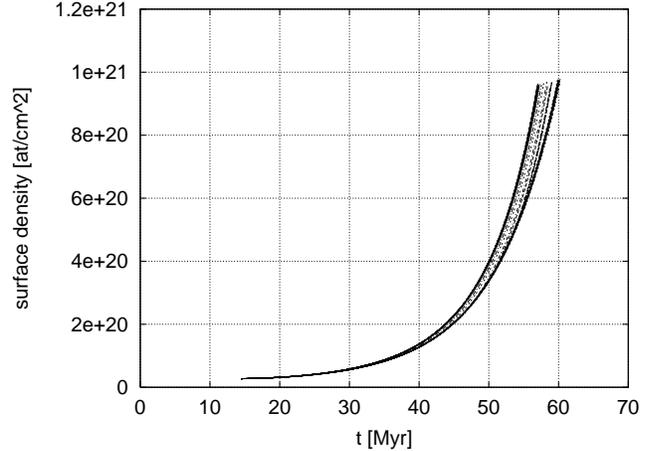}}
\caption{The evolution of the maximum perturbation of the surface density
for intermediate phase-shifts between initial values of the linear and 
nonlinear terms of perturbation.} 
\label{dens3}
\end{figure}

The time evolution of $\Sigma_0$ and $\Sigma_1$ are presented 
in Figs.~\ref{dens1}, \ref{dens2}, and  \ref{dens3} and the corresponding 
amplitudes of
real and imaginary parts of $\xi , \xi_+ , \xi_-$ for the first two cases 
in Figs.~\ref{ampl1} and \ref{ampl2}. 
We can distinguish two situations: the linear and non-linear parts of the
perturbation are in phase, so that they support each other,
which is seen in Figs.~\ref{dens1} and \ref{ampl1}, 
or they are in anti-phase, so that the  nonlinear 
contribution slows down the linear growth of perturbation, as it is visible in
Figs.~\ref{dens2} and \ref{ampl2}.
Contribution of the non-linear terms depends on the shape of the forming
fragments, i.e. on the value of the amplitude functions $\xi ,\xi_+ , \xi_- $.
Figs.~\ref{dens1} - \ref{ampl2} show the extreme cases of that contribution.
Intermediate cases, keeping the initial value of the 
perturbation in surface density at the
same level, $\Sigma_1 / \Sigma_0 = 0.05$, are given in 
Fig.~\ref{dens3}. 
We can also see in Fig.~\ref{dens1} that the maximum contribution 
of nonlinear terms to 
the value of the perturbed surface density, at the time when 
$\Sigma_0 = \Sigma_1$, is $\sim 25 \%$ of the linear value.  
  
All the curves start at the instability time $t_\mathrm{b}$:  functions 
$\Sigma_{\vec\eta}$, 
$\Sigma_{\vec\eta_{+}}$ and $\Sigma_{\vec\eta_{-}}$ grow,
although the absolute value of the appropriate amplitude functions 
$\xi$, $\xi_{+}$ and
$\xi_{-}$ descend during the short time after the instability begins. 
It is because
$\Sigma_{\vec\eta}$, $\Sigma_{\vec\eta_{+}}$ and $\Sigma_{\vec\eta_{-}}$ are 
connected to the amplitude functions through the eigenvector 
(\ref{eigenvect}),  whose $\Sigma $ part is always growing with time.

\begin{figure}
\resizebox{\hsize}{!}{\includegraphics{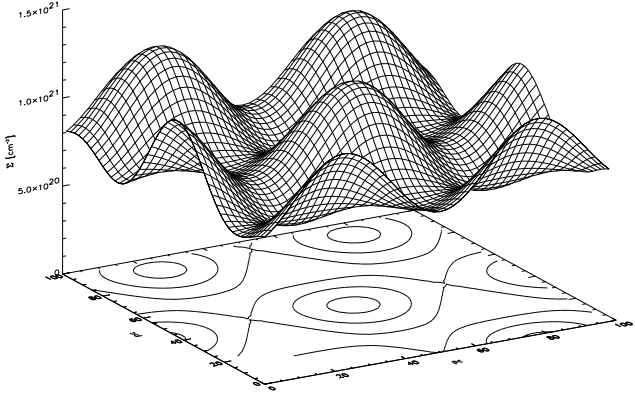}}
\caption{The distribution of $\Sigma $ in the tangential plane at the time
t = 55 Myr for the perturbation shown in Figs.~\ref{dens1} and \ref{ampl1}.} 
\label{denplane1}
\end{figure}

\begin{figure}
\resizebox{\hsize}{!}{\includegraphics{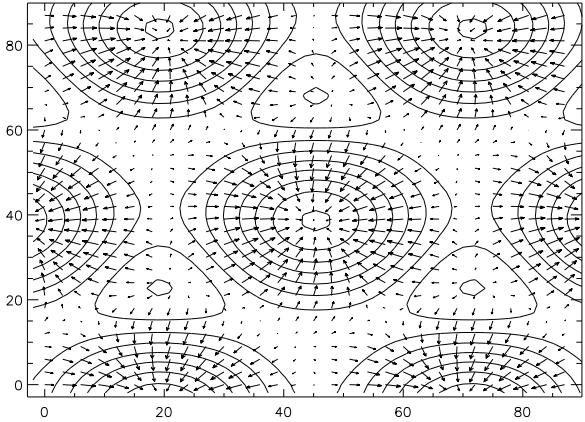}}
\caption{The velocity vectors of the surface flows $\vec v$ corresponding to 
Fig.~\ref{denplane1}.} 
\label{velplane1}
\end{figure}

\begin{figure}
\resizebox{\hsize}{!}{\includegraphics{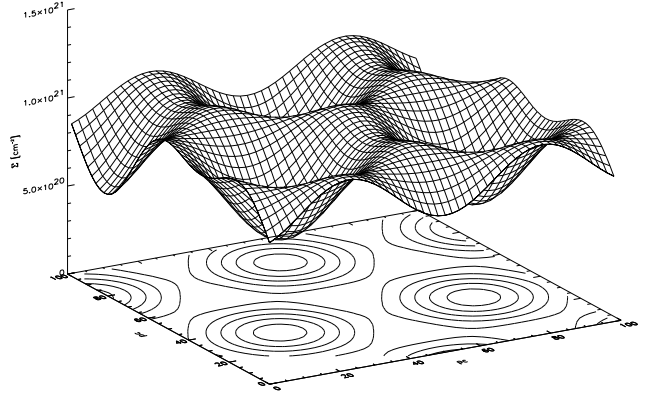}}
\caption{The same as in Fig.~\ref{denplane1} at the same time for the
perturbation shown in Figs.~\ref{dens2} and \ref{ampl2}.} 
\label{denplane2}
\end{figure}

\begin{figure}
\resizebox{\hsize}{!}{\includegraphics{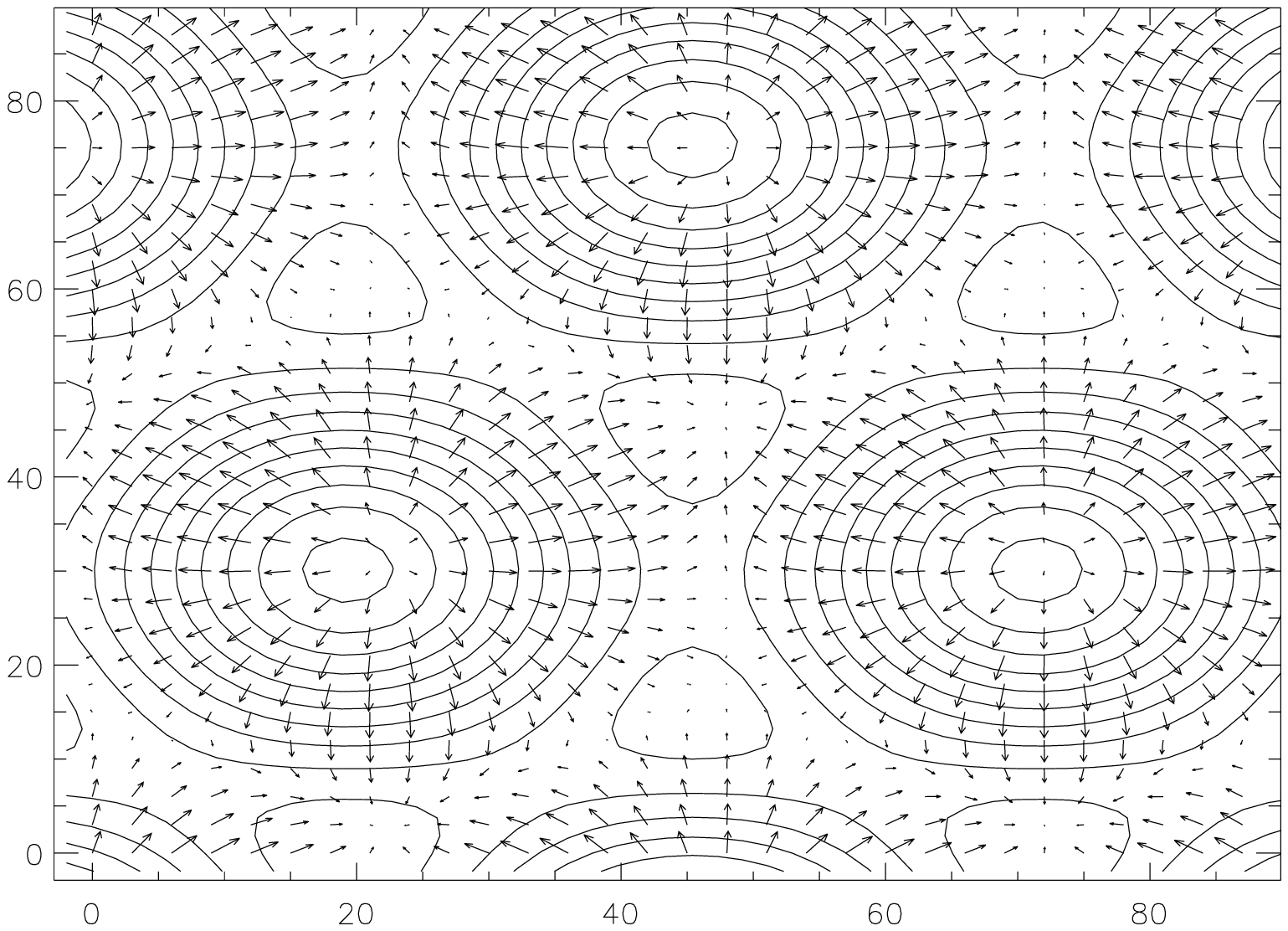}}
\caption{The velocity vectors of the surface flows $\vec v$ corresponding to 
Fig.~\ref{denplane2}.} 
\label{velplane2}
\end{figure}

The surface density $\Sigma $ at any point of the tangential plane 
may be computed at any expansion time after $t_\mathrm{b}$ using
eigenvectors (\ref{eigenvect}) and Eq. (\ref{fourier_t})
written for modes $(\vec\eta,\vec\eta_{+},\vec\eta_{-})$.
In Fig.~\ref{denplane1} we show the distribution of the 
surface density $\Sigma $ and in Fig.~\ref{velplane1} the velocity field of
the surface flows $\vec v$  
in the tangential plane for $\eta_\mathrm{max}$ in the case when the 
initial perturbations have 
linear terms in phase with the nonlinear terms corresponding to 
Figs.~\ref{dens1}
and \ref{ampl1} at the time $t = 55\ \mathrm{Myr}$. Figs.~\ref{denplane2} and 
\ref{velplane2} give $\Sigma $ and $\vec v$ in the tangential plane at the
same time for the case when the linear and nonlinear terms of the 
initial perturbation are in anti-phase corresponding to Figs.~\ref{dens2} and 
\ref{ampl2}. In the former case, the fragments are well defined and the 
density peaks are separated one from another with
deep depressions in $\Sigma $. In the later case, there are high 
surface density chains with no distinct peaks and we cannot separate 
individual fragments. 

\begin{figure}
\resizebox{\hsize}{!}{\includegraphics{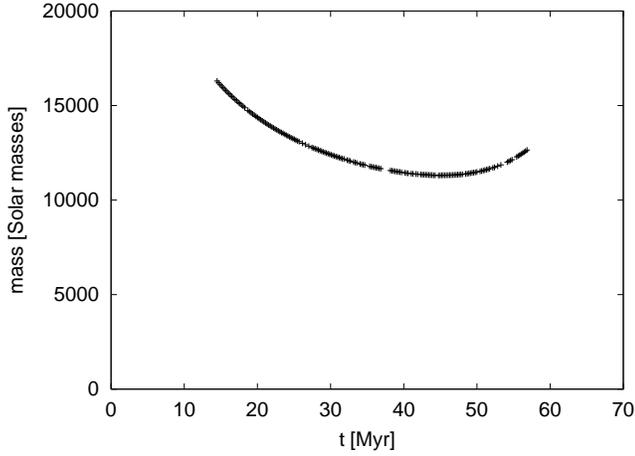}}
\caption{The time evolution of total mass in a well defined fragment.} 
\label{masstime}
\end{figure}

\begin{figure}
\resizebox{\hsize}{!}{\includegraphics{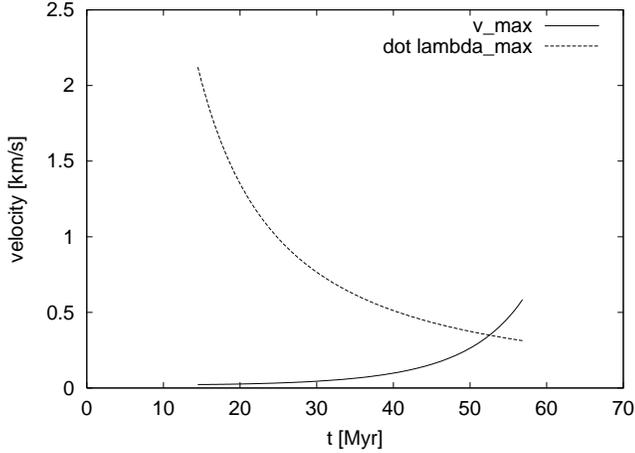}}
\caption{$\dot \lambda $ and $v_\mathrm{max}$ as functions of time.} 
\label{dlabdv}
\end{figure}

We measure 
the total mass concentrated in one of the well defined 
fragments shown in Fig.~\ref{denplane1}. The mass of this fragment  
is given 
in Fig.~\ref{masstime} as a function of time.
It decreases because the decrease of its size, which is proportional to 
$\lambda_\mathrm{max} = {2 \pi R \over \eta_\mathrm{max}}$, and increases because the 
accumulation of the ambient medium and surface flows. After $t_\mathrm{b}$ the  
resulting mass of the fragment 
decreases, since the influence of the size shrinking dominates.
This happens when the magnitude of ${\mathrm{d} \lambda_\mathrm{max} \over \mathrm{d}t}$ is larger 
then the amplitude of the surface flows $v_\mathrm{max}$ (see Fig.~\ref{dlabdv}).
The magnitude of ${\mathrm{d} \lambda_\mathrm{max} \over \mathrm{d}t}$ decreases with time and 
$v_\mathrm{max}$ increases, and at
$t \sim 53\ \mathrm{Myr}$ they are equal. Since then the inflow dominates and the 
fragment mass growths.    

\subsection{The mass spectrum of fragments}

\begin{figure}
\resizebox{\hsize}{!}{\includegraphics{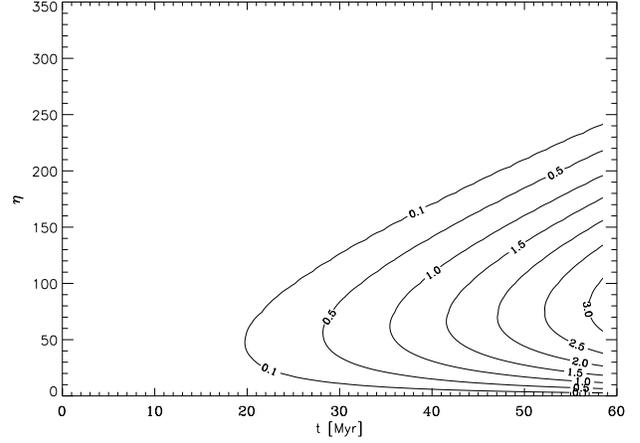}}
\caption{The time evolution of the fragmentation integral $I_{\mathrm{f}, \eta}$ as
given by Eq. (\ref{genfrag}).} 
\label{fragint}
\end{figure}

After $t_\mathrm{b}$, when the first mode begins to be gravitationally unstable,
more and more modes are unstable and the interval of instability growths.
In Fig.~\ref{fragint} we give the values of the fragmentation integral 
$I_{\mathrm{f}, \eta}(t)$ as defined in (\ref{genfrag}) as a function of time.
This shows at any time the level of development of a fragment with given
$\eta $.  

\begin{figure}
\resizebox{\hsize}{!}{\includegraphics{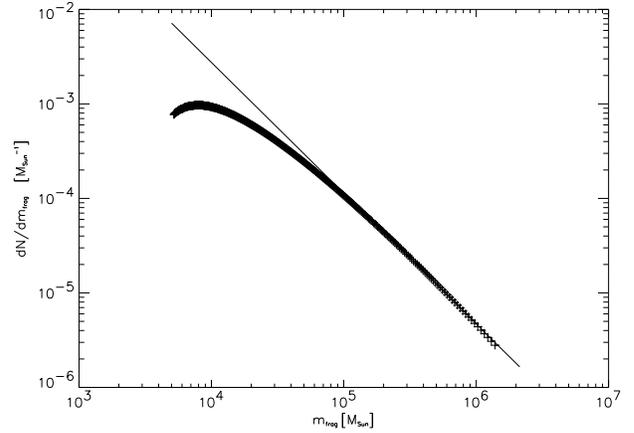}}
\caption{The mass spectrum of fragments. The straight line is the power law
fit of the decreasing part of the spectrum $m^{-1.4}$.} 
\label{massspec}
\end{figure}

\begin{figure}
\resizebox{\hsize}{!}{\includegraphics{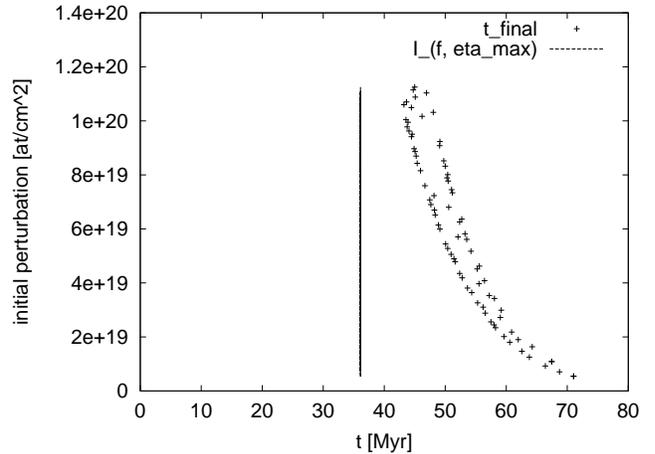}}
\caption{Dependence of the fragmentation time on the initial
perturbation of the surface density. The vertical line gives the time when 
the value of $I_{\mathrm{f}, \eta_\mathrm{max}} = 1$. } 
\label{tfrag}
\end{figure}

Mass $m_\mathrm{frag}$ of a fragment, which is related to 
$\lambda = {2 \pi R \over \eta}$, may be defined as 
\begin{equation}
m_\mathrm{frag} = \pi \left({\lambda \over 2}\right)^2 \Sigma_0 = 
{\pi ^3 R^2 \Sigma_0 \over \eta ^2}.
\label{masses}
\end{equation}
The formation frequency of fragments corresponding to the wavenumber $\eta $
is proportional to $I_{\mathrm{f}, \eta }$, and the surface of the spherical shell of 
radius
$R$ is able to accommodate $R^2/ \lambda^2$ fragments of wavelength $\lambda $.
Therefore, 
the number $\mathrm{d}N$ of fragments of mass $m_\mathrm{frag}$
formed out of the spherical shell of radius R in gravitational unstable modes
with wavenumber $\eta $ from the interval $(\eta, \eta + \mathrm{d}\eta)$ is
\begin{equation}
\mathrm{d}N = I_{\mathrm{f}, \eta} \times {R^2 \over \lambda ^2} \mathrm{d}\eta.
\end{equation}  
With this equation, we may derive the mass spectrum of fragments 
$\mathrm{d}N/\mathrm{d}m_\mathrm{frag}$  at some time.
It is shown in Fig.~\ref{massspec} at the time
$t_\mathrm{f}$, when $\Sigma_0 = \Sigma_1$. 
The masses are between 
a few times $10^3\ \mathrm{M_{\odot}}$ and $1.5 \times 10^6\ \mathrm{M_{\odot}}$ with the 
highest probability peak at $m_\mathrm{frag} \simeq 10^4\ \mathrm{M_{\odot}}$. 
At that time is
the radius of the shell almost $1 \mathrm{kpc} $ with the expansion velocity $\sim
5\ \mathrm{km\, s}^{-1}$, and the total mass collected in the shell is $1.14 \times 10^9
\ \mathrm{M_{\odot}}$.  

The decreasing part of the mass spectrum can be approximated
as a power law $\mathrm{d}N/\mathrm{d}m_\mathrm{frag} \sim m_\mathrm{frag}^{\alpha }$: 
the fit of the this part of $\mathrm{d}N/\mathrm{d}m_\mathrm{frag}$ gives 
$\alpha  = - 1.4$, which is close to the observed
mass spectrum of GMC in the Milky Way:  Combes(\cite{combes}) 
gives $\alpha = -1.5$.
NANTEN survey of the CO emission of the LMC (Fukui, 2001) gives steeper 
slope of $\alpha = -1.9$, 
which may be explained in the connection to higher level of random 
velocities in the LMC compared to the Milky Way 
resulting in the deficit of high mass clouds.

\subsection{The time of fragmentation}

The evolution of the maximum perturbation of the surface density
can be used to determine the fragmentation time $t_\mathrm{f}$ of the shell.
Because at advanced stages the value of the maximum 
perturbation rises steeply with the time (see e.g. Fig.~\ref{dens1}), we 
define $t_f$ as the time, when maximum perturbation of the surface density is 
equal to the unperturbed value: $\Sigma_1(t_\mathrm{f}) = \Sigma_0(t_\mathrm{f})$.
Using the fragmentation integral $I_{\mathrm{f}, \eta }$ we may compare the 
development level of different fragments at $t_\mathrm{f}$ (see Fig.~\ref{fragint}). 
We can say that the most frequent 
fragments are also the most developed, the more massive form only later.

$t_\mathrm{f}$ depends on the
initial conditions of the set of Eqs. (\ref{finalequ}). They correspond
to the initial perturbation of the surface density. We can set them
to the value typical for the inhomogeneities in the clumpy interstellar
medium ($10^{19}~-~10^{20}~\mathrm{cm}^{-2}$), which is at $t_\mathrm{b}$: $0.01 - 0.2 \times 
\Sigma_0$.

The dependence of $t_\mathrm{f}$ on the value of the initial perturbation, 
$\Sigma_1(t_\mathrm{b})$, is shown in 
Fig.~\ref{tfrag}. Fragments form since $45\ \mathrm{Myr}$, for the largest 
perturbations, to $70\ \mathrm{Myr}$, for the smallest  perturbations. 
The spread in $t_\mathrm{f}$ for given $\Sigma_1(t_\mathrm{b})$ 
is connected
to the different shape of the  perturbation as shown in 
Fig.~\ref{dens3}. This time may be compared to the fragmentation 
time $t_{\mathrm{f}, \eta }$ obtained for $\eta_\mathrm{max}$ from the linear analysis 
defined as a time when $I_{\mathrm{f}, \eta} = 1$.

\section{Conclusions}

We evaluate the time evolution of perturbations on the surface of an 
expanding shell. We complement  the linear analysis of the gravitational 
fragmentation process with the inclusion of 
nonlinear terms, and we compute the time evolution of fragments
after the time when the shell starts to be unstable.
Some initial perturbations develop into well separated fragments and we 
estimate the time evolution of the mass of a fragment, the mass 
spectrum of fragments, and the spread in their formation time. 
The computed mass spectrum is 
close to the observed mass distribution 
of  GMC in the Milky Way, but slightly flatter than the mass spectrum of 
molecular clouds observed in the LMC.
This may be related to higher level of random motions in the LMC compared
to the Milky Way, which
restricts the formation of late time massive 
fragments and steepen the resulting mass spectrum.  
Also interesting is that the more massive fragments 
form at later times of the shell evolution than the less massive 
fragments. The formation time 
depends on the value of the initial perturbation: $t_\mathrm{f} = 45 - 70\ \mathrm{Myr}$.
Large density fluctuations shorten this time and thus in the disturbed ISM
with large density fluctuations the fragments form sooner than in quiet 
and smooth ISM where the density fluctuations are small.

\begin{acknowledgements}

We would like to thank Burkhard Fuchs and to anonymous referee for valuable 
comments.
This work was inspired by the paper on the fragmentation of uniformly
rotating disks by Fuchs (\cite{fuchs}).
We are also grateful for an enlighting discussions with B.~Fuchs 
in April 1998 and
in March 2000 at Star~2000 conference in Heidelberg.
The authors gratefully acknowledge financial support by the Grant Agency 
of the Academy 
of Sciences of the Czech Republic under the grant No.~A~3003705/1997 and 
support by the 
grant project of the Academy of Sciences of the Czech Republic 
No.~K1048102. 
\end{acknowledgements}

\end{document}